# Optimizing Time-resolved Magneto-optical Kerr Effect for High-fidelity Magnetic Characterization

Yun Kim,[1] Dingbin Huang,[1] Deyuan Lyu,[2] Haoyue Sun,[3] Jian-Ping Wang,[2] Paul A. Crowell,[3] and Xiaojia Wang[1, a)]

[1]*Department of Mechanical Engineering, University of Minnesota, Minneapolis, Minnesota 55455, USA*
[2]*Department of Electrical and Computer Engineering, University of Minnesota, Minneapolis, Minnesota 55455, USA*
[3]*School of Physics and Astronomy, University of Minnesota, Minneapolis, Minnesota 55455, USA*

**Abstract:** Spintronics has emerged as a key technology for fast and non-volatile memory with great CMOS compatibility. As the building blocks for these cutting-edge devices, magnetic materials require precise characterization of their critical properties, such as the effective anisotropy field ($H_{k,eff}$, related to magnetic stability) and damping ($\alpha$, a key factor in device energy efficiency). Accurate measurements of these properties are essential for designing and fabricating high-performance spintronic devices. Among advanced metrology techniques, Time-resolved Magneto-Optical Kerr Effect (TR-MOKE) stands out for its superb temporal and spatial resolutions, surpassing traditional methods like ferromagnetic resonance (FMR). However, the full potential of TR-MOKE has not yet been fully pledged due to the lack of systematic optimization and robust operational guidelines. In this study, we address this gap by developing experimentally validated guidelines for optimizing TR-MOKE metrology across materials with perpendicular magnetic anisotropy (PMA) and in-plane magnetic anisotropy (IMA). Our work identifies the optimal ranges of the field angle to simultaneously achieve high signal amplitudes and improve measurement sensitivities to $H_{k,eff}$ and $\alpha$. By suppressing the influence of inhomogeneities and boosting sensitivity, our work significantly enhances TR-MOKE capability to extract magnetic properties with high accuracy and reliability. This optimization framework positions TR-MOKE as an indispensable tool for advancing spintronics, paving the way for energy-efficient and high-speed devices that will redefine the landscape of modern computing and memory technologies.

[a)]Author to whom correspondence should be addressed. Electronic mail: wang4940@umn.edu

Spintronics, leveraging both the charge and spin of electrons, offers promising advantages for information processing beyond conventional CMOS-based memories.[1,2] With high endurance, nonvolatility, and ultrafast switching speeds, devices such as spin-transfer torque magnetic random-access memories are driving innovation in memory and computing technologies.[1-4] Particularly, the advancements in cloud computing, 5G station, and artificial intelligence, accelerates the demand for energy-efficient and high-performance spintronic devices. Achieving this requires precise characterization of key magnetic properties, such as the effective anisotropy field ($H_{k,eff}$) and damping ($\alpha$), pivotal for ensuring data retention, low energy consumption, and fast dynamics.[2,5] Accurate determination and further manipulation of these properties are essential for optimizing material performance for spintronic applications.

Among advanced metrology techniques for studying magnetic materials, time-resolved magneto-optical Kerr effect (TR-MOKE) has exceled for its superb spatial and temporal resolutions, surpassing conventional methods like ferromagnetic resonance.[4,6-9] The high spatial resolution of TR-MOKE enables probing of diffraction-limited areas (down to ~sub-micron scales), which facilitates spatial mapping and the study of laser-induced spin-wave propagation.[10-12] While the latter may introduce complexities in analyzing inhomogeneous broadening, it also presents opportunities for exploring rich spin dynamics.[13] With sub-picosecond laser pulses, TR-MOKE can capture ultrafast phenomena such as phonon-magnon coupling,[14,15] optical switching,[3,16,17] ultrafast demagnetization,[3,18-20] and THz detection.[21,22] Additionally, TR-MOKE has proven effectiveness for probing interlayer exchange coupling in synthetic antiferromagnets,[23,24] making it essential for spintronic research.

Despite its advantages, the full potential of TR-MOKE is limited by the lack of a systematic understanding and optimization framework to further enhance its measurement accuracy and sensitivity. In an earlier work, Lattery *et al.* reported the signal amplitudes optimization for better signal-to-noise ratios (SNRs);[25] however, studies of measurement sensitivities remain elusive. Here, we propose a comprehensive optimization framework for the TR-MOKE metrology to simultaneously achieve high signal amplitudes and sensitivity. This optimization framework is developed for two practical scenarios of polar TR-MOKE measurements: (1) sweeping the magnetic field strength ($H_{ext}$) at a fixed field angle [$\theta_H$, defined as the angle between the field direction and the sample surface normal ($z$-axis), see Fig. 1(b)], and (2) sweeping $\theta_H$ while keeping $H_{ext}$ constant.[8] The proposed optimization of measurement conditions is validated through comparisons between optimization calculations and TR-MOKE measurements. Since our TR-MOKE setup is operated in scenario 1, the validation measurements are conducted by sweeping $H_{ext}$ at a few discrete $\theta_H$ on a $Co_{20}Fe_{60}B_{20}$ (CoFeB) thin film with perpendicular magnetic anisotropy (PMA).

In polar TR-MOKE measurements, the signal amplitude is proportional to the change in the $z$-component of magnetization ($\Delta M_z$).[4,6,25-27] $\Delta M_z$ arises from the instantaneous decrease in $H_{k,eff}$ induced by ultrafast laser heating (referred to as thermal demagnetization, in which optically excited 'hot' electrons pass thermal energy and angular momentum to other reservoirs such as spins and lattice).[18,19] The rapid change alters the equilibrium position of magnetization, represented by the change in the polar angle from $\theta_1$ to $\theta_2$ in Fig. 1(b), initiating magnetization precession. As the system cools down, $H_{k,eff}$ experiences a recovery process (also referred to as remagnetization).[3,19,28] In the optimization calculation, thermal demagnetization process is treated as a 5% step decrease in $H_{k,eff}$, persisting for 1.50 ps ($\Delta t$) before a step recovery to 99% of its initial value [Fig.1(a)]. The choices of these values are based on literature data[19] and pulse duration measurements.[23] Note that our final conclusion is not impacted by these initial values in the optimization model. Assuming the amplitude of magnetization ($|\mathbf{M}|$) remains constant, the signal amplitude can be expressed by:

$$\text{Amp} \propto \sqrt{R^2 + r^2 - 2rR \cos(2\pi f \Delta t)} \sin \theta_3 \qquad (1)$$

where $f$ is the precession frequency, $R = \theta_1 - \theta_2$, and $r = \theta_3 - \theta_2$ with $\theta_2$ and $\theta_3$ being the polar angles of the equilibrium positions corresponding to the 5% step decrease in $H_{\text{k,eff}}$ and the subsequent recovery, respectively, as denoted in Fig.1(c). Further details on the amplitude model are provided in Section S1 of the Supplementary Material (SM).

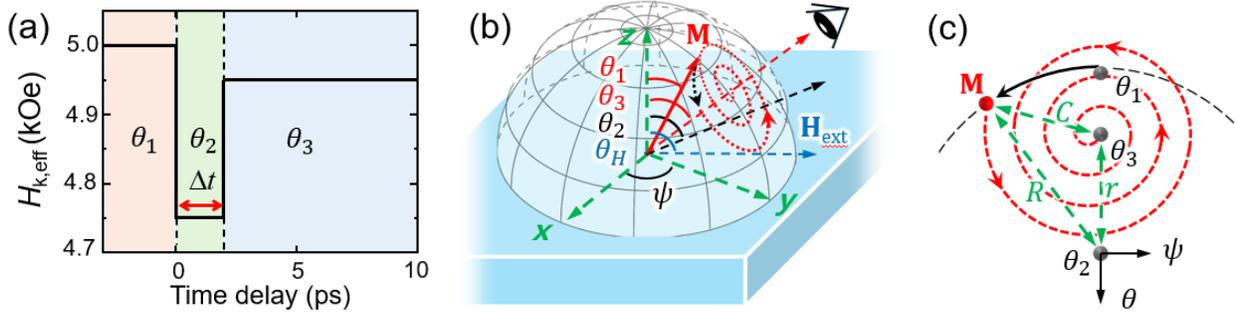

**FIG. 1.** (a) $H_{\text{k,eff}}$ profile as a function of time. $H_{\text{k,eff}}$ undergoes a 5% step decrease, lasting for 1.50 ps before it recovers to 99% of its original value. (b) Schematic representation of the parameters and physical variables used in the model calculations, depicted in the global spherical coordinates. (c) Magnetization dynamics following the $H_{\text{k,eff}}$ profile in (a). The parameters of $R$ and $r$, used for calculating signal amplitudes, are illustrated in a local coordinate system projected onto the spherical surface viewed against the direction of $\theta_3$.

In addition to enhancing signal amplitudes, achieving high measurement sensitivities to the key properties (*e.g.*, anisotropy and damping) is also essential for the optimization. In TR-MOKE measurements, analysis of the field-dependent precession frequency enables the determination of $H_{\text{k,eff}}$ and $\theta_H$. The measurement sensitivities to $H_{\text{k,eff}}$ ($S_{H_{\text{k,eff}}}$) and $\theta_H$ ($S_{\theta_H}$) are defined as $S_{H_{\text{k,eff}}} = d\ln(f) / d\ln(H_{\text{k,eff}})$ and $S_{\theta_H} = d\ln(f) / d\ln(\theta_H)$, respectively. These sensitivities are expressed using natural logarithms to enable direct propagation of relative uncertainties during error analysis.[29] Additionally, the relaxation rate ($1/\tau_{\text{tot}}$) analysis provides insight into the damping and inhomogeneous broadening contributions from sources such as variations in $H_{\text{k,eff}}$ or easy-axis orientations within the probed area. The relaxation rate can be expressed as the summation of these contributions:[6,7,26]

$$\frac{1}{\tau_{\text{tot}}} = \frac{1}{\tau_\alpha} + \frac{1}{\tau_{\Delta H_{\text{k,eff}}}} + \frac{1}{\tau_{\Delta\theta_H}} = \frac{1}{2}\alpha\gamma(H_1 + H_2) + \pi\left|\frac{df}{dH_{\text{k,eff}}}\right|\Delta H_{\text{k,eff}} + \pi\left|\frac{df}{d\theta_H}\right|\Delta\theta_H \qquad (2)$$

where $1/\tau_\alpha$, $1/\tau_{\Delta H_{\text{k,eff}}}$, and $1/\tau_{\Delta\theta_H}$ represent the relaxation rate contributions from damping, the inhomogeneous broadening of $H_{\text{k,eff}}$ ($\Delta H_{\text{k,eff}}$), and that of the easy axis (translated to $\Delta\theta_H$). $\gamma$ denotes the gyromagnetic ratio, $H_1 = H_{\text{ext}}\cos(\theta - \theta_H) + H_{\text{k,eff}}\cos^2(\theta)$, and $H_2 = H_{\text{ext}}\cos(\theta - \theta_H) - H_{\text{k,eff}}\cos(2\theta)$.[26,30] Sensitivities to each property can be calculated as the fractional contribution of the respective relaxation rate to the total relaxation rate. For example, sensitivity to $\alpha$ ($S_\alpha$) is calculated as $S_\alpha = (1/\tau_\alpha)/(1/\tau_{\text{tot}})$.

To determine the optimal conditions for maximizing signal amplitudes and sensitivities to specific properties, two objective functions are defined: one for amplitude (Amp) and the other for sensitivity (Amp·$S_X$) where $S_X$ is the measurement sensitivity to a specific property of interest. The optimal field angle is identified by maximizing the objective function:

$$\theta_{\text{opt}}^{\text{Amp}} = \theta_H[\max(\text{Amp})] \qquad (3)$$

and

$$\theta_{\text{opt}}^{S_X} = \theta_H[\max(\text{Amp} \cdot S_X)] \qquad (4)$$

Here, $\theta_{\text{opt}}^{\text{Amp}}$ and $\theta_{\text{opt}}^{S_X}$ represent the optimal field angles for maximizing Amp and Amp·$S_X$, respectively, at each $H_{\text{ext}}$. Given similar levels of noise floor throughout all experiments, higher signal amplitudes collected at $\theta_{\text{opt}}^{\text{Amp}}$ can improve the data quality, leading to more accurate determination of $f$ and $\tau$. Since the uncertainties of $f$ and $\tau$ propagate to the uncertainties of the extracted magnetic properties, it is necessary to consider both signal amplitude[31] and sensitivity when identifying $\theta_{\text{opt}}^{S_X}$.

The normalized TR-MOKE signal amplitudes for PMA and IMA materials under varying $H_{\text{ext}}$ and $\theta_H$ are calculated based on Eq. (1). The results are plotted in Figs. 2(a) and 2(d) as contours of the normalized external field ($H_{\text{ext}}/H_{\text{k,eff}}$) from 0 to 10 and $\theta_H$ from 0° to 90°. For PMA materials, $\theta_{\text{opt}}^{\text{Amp}}$ remains at 90° when $H_{\text{ext}}/H_{\text{k,eff}} < 1$ and approaches ~60° as $H_{\text{ext}}/H_{\text{k,eff}}$ increases. When $H_{\text{ext}}/H_{\text{k,eff}} < 1$, magnetic anisotropy energy dominates magnetic free energy. In this regime, the difference in equilibrium position before and after heating ($R = \theta_1 - \theta_2$) becomes larger as $\theta_H$ approaches 90°, leading to the highest amplitude occurring at 90°. For $H_{\text{ext}}/H_{\text{k,eff}} > 1$ where the Zeeman energy dominates, the difference in equilibrium position is zero when $\theta_H = 90°$ since **M** is saturated along the in-plane direction, resulting no initial torque for initiating precession. To get the large signal amplitudes in this regime, $\theta_H$ should deviate from 90°, approaching ~60°. This trend of signal amplitudes on $\theta_H$ is also observed from validation measurements of a reference CoFeB film, indicated by the white dots in Fig. 2(a) to be further discussed later. As for IMA, $\theta_{\text{opt}}^{\text{Amp}}$ is 0° when $H_{\text{ext}}/H_{\text{k,eff}} < 1$ and converges to ~50° as $H_{\text{ext}}/H_{\text{k,eff}}$ exceeds 1. However, $\theta_H = 0°$ is not recommended, as achieving high amplitudes at this angle requires low fields, which can result in incoherent spin procession.

Considering our TR-MOKE setup is a sweeping-field configuration at a fixed $\theta_H$ (scenario 1), the optimal field angle for practical TR-MOKE measurements shall be recommended based on the range of applied fields, instead of individual $H_{\text{ext}}$. For example, for samples with low $H_{\text{k,eff}}$, it is relatively easier to reach a high ratio of $H_{\text{ext}}/H_{\text{k,eff}}$; while for samples with high $H_{\text{k,eff}}$, achieving a high ratio of $H_{\text{ext}}/H_{\text{k,eff}}$ becomes challenging with the same maximum $H_{\text{ext}}$. To encompass

sample- and $H_{ext}$-dependent variations, we define an average optimal angle ($\bar{\theta}_{opt}^{Amp}$) as the angle that yields the maximum average amplitude over a specific range of $H_{ext}/H_{k,eff}$ for a given $\theta_H$. For various ranges of $H_{ext}/H_{k,eff}$ with the lower limit from 0 to 3 and the upper limit from 4 to 6, the recommended $\bar{\theta}_{opt}^{Amp}$ are 61° - 79° for PMA and 32° - 50° for IMA materials, respectively (see Fig. S4 in Sec. S2 of the SM for more details).

The calculated sensitivities ($\log_{10}$ is taken after normalization to the global maximum) and the corresponding optimal angle of $\theta_{opt}^{S_X}$ [defined in Eq. (4)] are shown in Fig. 2. For $S_{H_{k,eff}}$, the optimization model suggests $\theta_{opt}^{S_{H_{k,eff}}}$ is 90° for PMA materials when $H_{ext}/H_{k,eff} < 1$ and gradually decreases to ~75° when $H_{ext}/H_{k,eff} > 1$ [Fig. 2(b)]. For IMA materials, $\theta_{opt}^{S_{H_{k,eff}}}$ is 0° when $H_{ext}/H_{k,eff} < 1$ and approaches ~30° when $H_{ext}/H_{k,eff} > 1$ [Fig. 2(e)]. To account for the amplitude error propagation to sensitivity, $\bar{\theta}_{opt}^{S_X}$ is chosen by minimizing the objective function of $1/\sqrt{\Sigma_{H_{ext}}(\text{Amp} \cdot S_X)^2}$ over the field range. For PMA materials, $\bar{\theta}_{opt}^{S_{H_{k,eff}}}$ ranges from 43° to 49° for $H_{ext}$ starting from a relatively low field (*i.e.*, $H_{ext} < 2H_{k,eff}$) and from 80° to 89° for $H_{ext}$ starting from a high field (*i.e.*, $H_{ext} > 2H_{k,eff}$). These ranges reflect the practical limitations of the achievable field range in TR-MOKE measurements. A lower starting field is beneficial for materials with larger $H_{k,eff}$, while a higher starting field is favored for materials with smaller $H_{k,eff}$, ensuring the collection of adequate data points across the available field range. $\bar{\theta}_{opt}^{S_{H_{k,eff}}}$ falls between 1° and 23° for IMA materials. Regarding the sensitivity to the easy axis ($S_{\theta_H}$), $\theta_{opt}^{S_{\theta_H}}$ [Figs. 2(c) and 2(f)] follows the trend of $\theta_{opt}^{Amp}$. $\bar{\theta}_{opt}^{S_{\theta_H}}$ varies from 66° to 89° for PMA materials and from 27° to 48° for IMA materials.

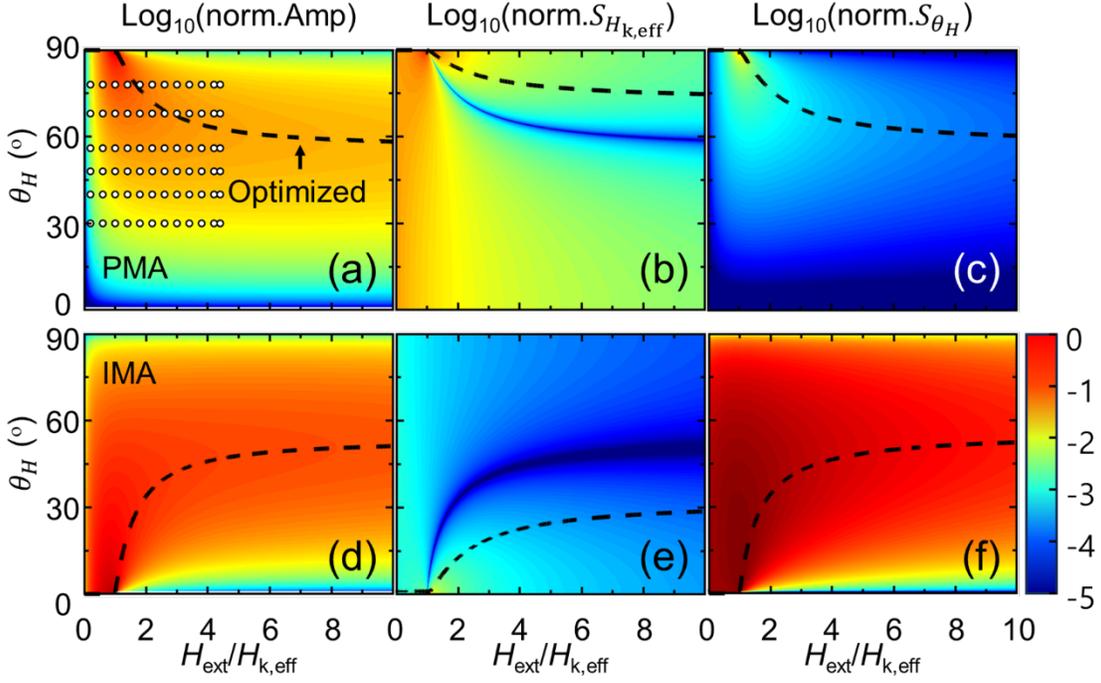

**FIG. 2.** Contour plots of the normalized amplitudes [(a) and (d)], sensitivities to $H_{k,eff}$ [(b) and (e)], and sensitivities to $\theta_H$ [(c) and (f)] of PMA [(a)-(c)] and IMA [(d)-(f)] materials under varying $H_{ext}/H_{k,eff}$ and $\theta_H$ (all plots are in the logarithmic scale). The white dots in (a) represent the signal amplitudes obtained from validation measurements taken at each $H_{ext}/H_{k,eff}$ for six discrete angles of $\theta_H$ (to be further discussed later). Black dashed curves represent $\theta_{opt}^{Amp}$ and $\theta_{opt}^{S_X}$ as functions of $H_{ext}/H_{k,eff}$ for the highest amplitude and maximum value of the objective function. Properties used in the optimization calculation are $H_{k,eff} = \pm 5.00$ kOe, $\gamma = 18.00$ rad/kOe/ns, and $\alpha = 0.020$, close to the magnetic properties of the CoFeB reference sample used in experimental validation. Ticks are negative as data are normalized and taken $\log_{10}$.

The relaxation-time-based sensitivities to the damping ($S_\alpha$) and inhomogeneous broadening due to $H_{k,eff}$ ($S_{\Delta H_{k,eff}}$) are calculated and plotted in Fig. 3. The reference sample (CoFeB thin film) for model validation exhibits uniaxial interface anisotropy and negligible $\Delta\theta_H$ contribution. Thus, the sensitivity to $\Delta\theta_H$ is omitted here (see Section S3 of the SM). . As shown in Fig. 3(a), for PMA materials, $\theta_{opt}^{S_\alpha}$ increases from ~55° to 90° when $H_{ext}/H_{k,eff} < 1$, and then converges to ~60° (dashed lines). For sweeping-field measurements with $\theta_H$ fixed, the average $\bar{\theta}_{opt}^{S_\alpha}$ spans from 62° to 74° over the range of $H_{ext}/H_{k,eff}$ (see Fig. S5 in the SM for more details). $S_{\Delta H_{k,eff}}$ is high when $H_{ext}/H_{k,eff} < 1$ and decreases as $H_{ext}$ increases, as shown in Fig. 3(b). The trend for minimizing $S_{\Delta H_{k,eff}}$ aligns with that of optimizing $S_\alpha$ when $H_{ext}/H_{k,eff} > 1$, since maximizing $S_\alpha$ automatically reduces $S_{\Delta H_{k,eff}}$. For IMA materials, $\bar{\theta}_{opt}^{S_\alpha}$ falls between 36° and 48°, where $S_{\Delta H_{k,eff}}$ is also minimized (see Fig. S5 in the SM for more details).

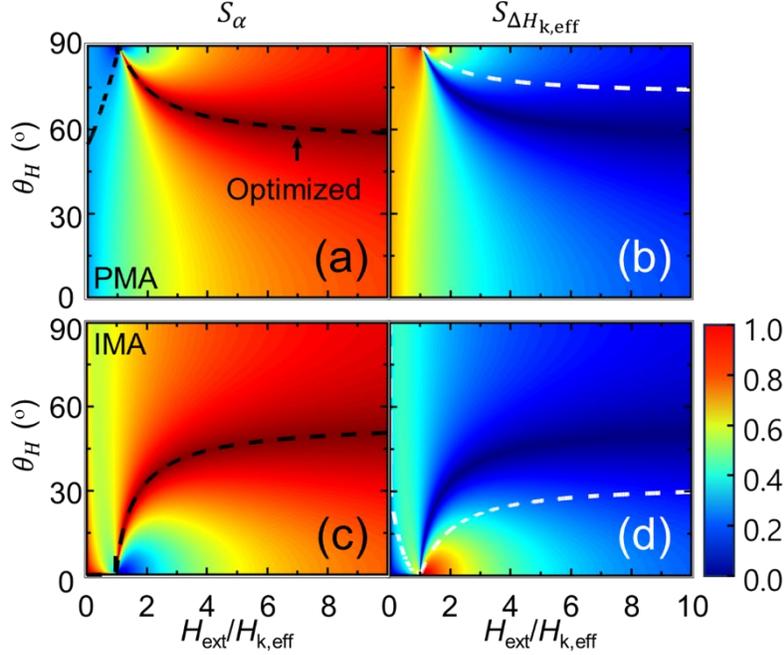

**FIG. 3.** Contour plots of sensitivity to $\alpha$ [(a) and (c)] and $\Delta H_{k,eff}$ [(b) and (d)] of PMA [(a) and (b)] and IMA [(c) and (d)] materials under varying $H_{ext}/H_{k,eff}$ and $\theta_H$. Black and white dashed curves represent the recommended $\theta_{opt}^{S_x}$ at each $H_{ext}/H_{k,eff}$ to achieve optimal sensitivity.

To validate the optimization framework, a series of TR-MOKE measurements are conducted on a CoFeB reference sample under $H_{ext}$ ranging from 3 to 22 kOe. Our TR-MOKE setup utilizes a femtosecond laser with an 80-MHz repetition rate and a center wavelength of 783 nm. The power is set as 8 mW for the pump beam modulated at 9 MHz and 4 mW for the probe beam modulated at 200 Hz. Both beams are focused onto the sample surface *via* a 10× objective, resulting a $1/e^2$ beam size of 6 μm. To achieve a wider tuning range of $\theta_H$, we employ a custom-designed sample holder in the Voigt geometry[32] and a compensator to correct phase changes introduced by the additional mirrors in the sample holder (see Section S4 in the SM for details). The sample stack consists of Si/SiO$_2$(sub)/W(5)/CoFeB(1.3)/MgO(2)/Ta(3) where numbers in parentheses are thickness in nm [Fig. 4(a)]. The TR-MOKE signal is fitted to Kerr angle, $\Delta\theta_K = A + Be^{-t/C} + \text{Amp}\sin(2\pi ft + \varphi)e^{-t/\tau}$, to extract the precession frequency ($f$), relaxation rate ($1/\tau$), and precession amplitude (Amp).[26] Figure 4(b) depicts the representative TR-MOKE signals and the corresponding best-fits (black curves). $H_{k,\text{eff}}$ is determined to be $5.00 \pm 0.13$ kOe by TR-MOKE, which agrees well with the value reported in previous literatures.[26,33] $\gamma$ and $\alpha$ are determined to be $17.40 \pm 0.12$ rad/kOe/ns and $0.020 \pm 0.004$, respectively, at 78°. More details about the TR-MOKE metrology and data reduction can be found in Section S5 in the SM and prior studies.[4,6,14,23-26,34,35]

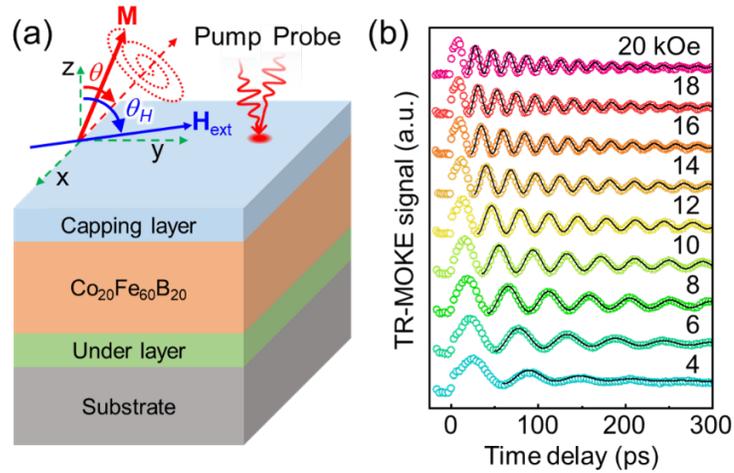

**FIG. 4.** (a) Illustration of the samples stack and physical parameters. (b) The representative TR-MOKE signals (symbols) and their best-fit curves (black lines) when $\theta_H = 70°$.

Figure 5(a) compares the normalized amplitudes from both optimization calculations and TR-MOKE measurements, as a function of $H_{\text{ext}}/H_{k,\text{eff}}$ for several discrete field angles [corresponding to white dots in Fig. 2(a)]. Here, $\theta_H$ of 30º and 40º are fixed in data reduction of TR-MOKE since they exhibit low sensitivity to $\theta_H$ at those angles [see Fig. 2(c)]. Generally, the maximum amplitude increases as $\theta_H$ varies from 30º to 78º. For cases with smaller $\theta_H$ (*i.e.*, $\theta_H < 50$º), the amplitude saturates with increasing $H_{ext}$. In contrast, for cases with larger $\theta_H$ (*i.e.*, 68º and 78º), the maximum amplitude occurs at $H_{ext}/H_{k,eff} \approx 1$ - 2, after which it decreases with $H_{ext}$. The measured amplitudes [symbols in Fig. 5(a)] align well with the calculated amplitudes, confirming the amplitude model. Figure 5(b) presents the normalized mean amplitudes with respect to $\theta_H$, averaged over the range of 7 kOe < $H_{ext}$ < 22 kOe. The experimental results are scaled to the calculated results to enable a clear trend comparison. The calculation (black curve) matches closely with experiment results (symbols), suggesting the maximum amplitude occurs at ~70º. The measurement uncertainties of $\theta_H$ are depicted as horizontal error bars in Fig. 5(b). Notably, larger deviations are only observed at smaller field angles, attributed to the low measurement sensitivity to $\theta_H$ when $\theta_H$ is small.

Figures 5(c) and 5(d) compare the uncertainties of $H_{k,eff}$ and $\theta_H$ from the optimization calculation and measurements, respectively. The uncertainty of a property from calculation ($\sigma_X$) is proportional to the objective function of $1/\sqrt{\Sigma_{H_{ext}}(\text{Amp} \cdot S_X)^2}$. For instance, a higher summation of the signal amplitude and sensitivity to the property of interest over a given $H_{ext}$ range leads to a smaller uncertainty in calculation. With a fixed noise level in TR-MOKE measurements, a higher signal amplitude improves SNRs and reduces the uncertainty of $f$. Similarly, the relative uncertainty of $H_{k,eff}$ can be estimated as $dH_{k,eff}/H_{k,eff} \propto 1/(\text{Amp} \cdot S_{H_{k,eff}})$. As shown in Fig. 5(c), the calculated the uncertainty of $H_{k,eff}$ decreases as $\theta_H$ increases, which reasonably captures the general trends of measurement results. The calculated uncertainty of a property goes to infinity due to zero amplitude at 90°, being impossible to conduct measurements. In Fig. 5(d), both the calculated and measured uncertainties of $\theta_H$ exhibit a decreasing trend as $\theta_H$ becomes larger. This consistency demonstrates that the optimization calculation provides a reasonably accurate prediction to tailor the uncertainties of $H_{k,\text{eff}}$ and $\theta_H$ expected from measurements. As for $\alpha$ and $\Delta H_{k,\text{eff}}$, further refinement of the current relaxation model [based on Eq. (2)] is required to optimize uncertainties to the damping and inhomogeneous broadening.[26]

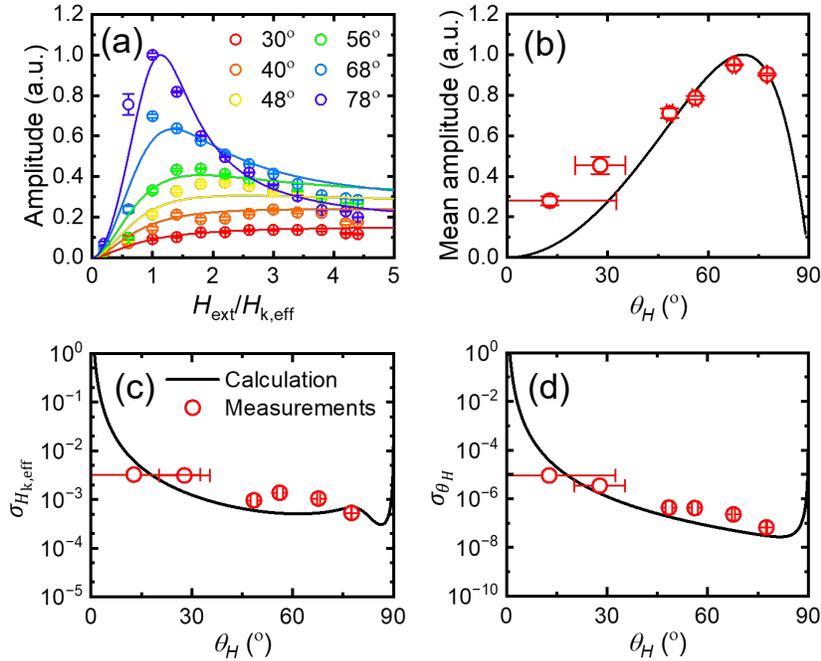

**FIG. 5.** Comparison between calculation and measurements. (a) Normalized signal amplitudes with respect to $H_{\text{ext}}/H_{k,\text{eff}}$ for several discrete field angles, corresponding to those white dots in Fig. 2(a). (b) Normalized mean amplitudes with respect to $\theta_H$, averaged over the range of 7 kOe < $H_{\text{ext}}$ < 22 kOe. The symbols are the average signal amplitudes of TR-MOKE measurements taken at individual external fields for a given $\theta_H$. Normalized uncertainty of $H_{k,\text{eff}}$ (c) and $\theta_H$ (d) as a function of $\theta_H$. Uncertainty from the calculation is evaluated in the $H_{\text{ext}}/H_{k,\text{eff}}$ range from 1.4 to 4.4, corresponding to the $f$ vs. $H_{\text{ext}}$ fitting range for experiments.

In summary, this study identifies the optimal field angle range for simultaneously achieving high signal amplitudes and enhanced measurement sensitivities to magnetic properties ($H_{k,\text{eff}}$, $\theta_H$, and $\alpha$) through a systematic optimization calculation and/or experimental validation. The findings provide recommended ranges of $\theta_H$ for studying PMA materials under the sweeping-field configuration (scenario 1): from 61º to 79º for maximizing signal amplitudes, from 43º to 49º or from 80º to 89º for extracting $H_{k,eff}$, and from 66º to 89º for determining $\theta_H$. For relaxation-time-based properties such as $\alpha$, the recommended $\theta_H$ range is from 62º to 74º, which also minimizes the influence of inhomogeneous broadening. For scenario 2 (sweeping-angle configuration), $H_{ext}$ around $2H_{k,eff}$ is suggested to balance signal amplitudes and sensitivity to $H_{k,eff}$ and $\theta_H$ for both PMA and IMA materials. $H_{ext} \approx 2H_{k,eff}$ is also recommended for the determination of $\alpha$ with suppressed inhomogeneous broadening (see Section S6 in the SM for details). This systematic optimization of TR-MOKE provides a clear pathway for more accurate and reliable measurements of material properties, critical for materials innovation in spintronic applications. Additionally, the improved SNRs *via* optimization could potentially enable the discovery of deeper insights into the rich underlying physics of magnetic materials.

## Author Declarations

The authors have no conflicts to disclose.

## Acknowledgments

This work is supported by the National Science Foundation (NSF) under award #2226579. Portions of this work were conducted in the Characterization Facility which receives partial support from the NSF through the MRSEC program (DMR-2011401), and the Minnesota Nano Center which is supported by the NSF through the National Nanotechnology Coordinated Infrastructure (ECCS-2025124). This article has been submitted to Applied Physics Letters. After it is published, it will be found at Link.

**Supplementary Material**

See Supplementary Material for details of the amplitude model, the identification of optimized field angle range, the effect of $\Delta\theta_H$, the amplitude compensation, and the scenario for sweeping-angle configuration.

**Data Availability Statement**

The data that support the findings of this study are available within the article or from the corresponding author upon reasonable request.

# Supplementary Material for "Optimizing Time-resolved Magneto-optical Kerr Effect for High-fidelity Magnetic Characterization"

Yun Kim,[1] Dingbin Huang,[1] Deyuan Lyu,[2] Haoyue Sun,[3] Jian-Ping Wang,[2] Paul A. Crowell,[3] and Xiaojia Wang[1, a)]

[1]*Department of Mechanical Engineering, University of Minnesota, Minneapolis, Minnesota 55455, USA*
[2]*Department of Electrical and Computer Engineering, University of Minnesota, Minneapolis, Minnesota 55455, USA*
[3]*School of Physics and Astronomy, University of Minnesota, Minneapolis, Minnesota 55455, USA*

## S1. Amplitude model with the step function of $H_{k,eff}$

Upon excitation by an ultrafast laser pulse, electrons in the material are the first to absorb energy, leading to a rapid and sharp increase in temperature which causes a reduction in magnetic anisotropy and thus the effective anisotropy field ($H_{k,eff}$). As the system evolves, the thermal energy and angular momentum of the electron subsystem will be transferred to other thermodynamic reservoirs (*i.e.*, magnetization and lattice subsystems).[1-3] As thermal equilibrium is restored between different subsystems, the temperature decreases and $H_{k,eff}$ starts to recover. Previous studies have shown that the energy and angular momentum transfer from electrons to other reservoirs occurs within ~2 ps, while the full recovery of $H_{k,eff}$ may take tens to hundreds of picoseconds.[1,4,5] Since $H_{k,eff}$ responds inversely to the transient temperature profile, we model this behavior as follows: $H_{k,eff}$ is set as 5.00 kOe prior to ultrafast laser excitation. Immediately after the pulse heating, $H_{k,eff}$ decreases by 5% to $0.95H_{k,eff}$, and remains at this level for 1.50 ps ($\Delta t$) as determined by the time scale of the laser excitation.[6] Following this, $H_{k,eff}$ recovers to $0.99H_{k,eff}$ and remains constant after $\Delta t$ [see Fig. 1(a)]. Note that the choice of these values are based on the

a)Author to whom correspondence should be addressed. Electronic mail: wang4940@umn.edu

literature values[1] and our conclusion is not impacted significantly by these values for the amplitude model.

In terms of magnetization dynamics, Fig. S1(a) illustrates the precession behavior following the ultrafast excitation. At the initial state, the magnetization vector (**M**) aligns to its initial equilibrium position, $\theta_1$. Upon excitation, the reduction in $H_{k,eff}$ shifts the equilibrium position to $\theta_2$, pulled closer to the direction of the external magnetic field $H_{ext}$, a process known as thermal demagnetization.[7] This abrupt shift creates a misalignment of **M** with its new equilibrium position, generating a torque that initiates precession around $\theta_2$ during the time interval of $\Delta t$. The extent of this magnetization precession corresponds to an angle of $2\pi f \Delta t$ [Fig. S1(b)]. After $\Delta t$, the equilibrium position changes from $\theta_2$ to $\theta_3$ as $H_{k,eff}$ recovers (*a.k.a.*, remagnetization). Consequently, **M** begins to precess around $\theta_3$ [Figs. S1(c) and S1(d)], following a new trajectory depicted as a red spiral curve.

The calculation of the signal amplitude requires the determination of $R$ and $r$, which are angles between equilibrium positions, as illustrated in Fig. S1(d). Considering the angular displacements on a spherical surface, $R = \theta_1 - \theta_2$ and $r = \theta_3 - \theta_2$. The angle between $R$ and $r$ corresponds to the precession angle during the ultrafast heating window of $2\pi f \Delta t$, which results in a deflection in the trajectory with $C$ being $\sqrt{R^2 + r^2 - 2rR\cos(2\pi f \Delta t)}$. As the TR-MOKE signal amplitude is proportional to the change in the $z$-component of magnetization ($\Delta M_z$), the signal amplitude can be expressed as Amp $\propto C\sin(\theta_3)$.

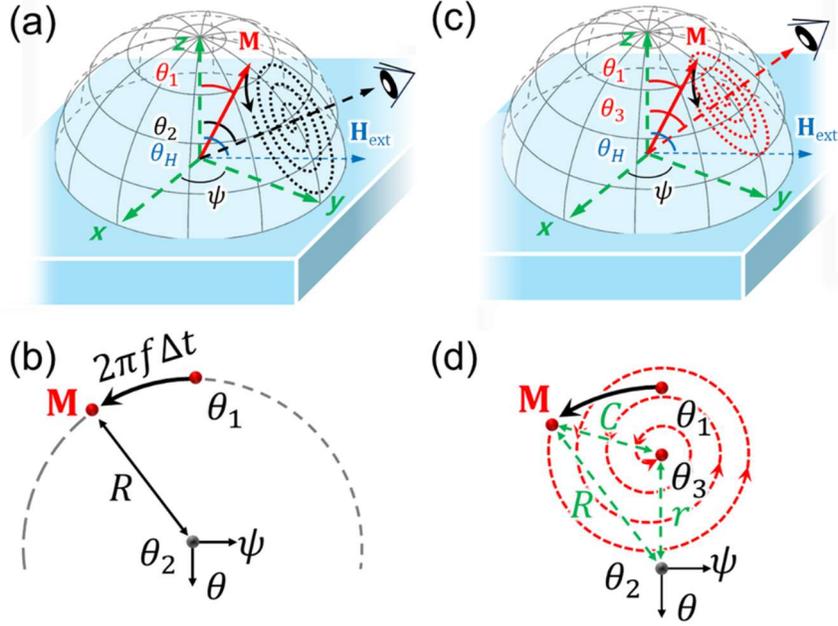

**FIG. S1.** (a) Illustration of magnetization dynamics after ultrafast heating. (b) The dynamics of **M** upon excitation and during the time interval of $\Delta t$. The trajectory of **M** from the viewpoint of against $H_{\text{eff}}$ is denoted as the dashed black line. (c) The dynamics of **M** after $\Delta t$, and the trajectory of **M** viewed against $H_{\text{eff}}$ denoted with dashed red line (d).

To ensure the generality of this work across samples with different thermal properties, we have examined how variations in these three parameters affect the amplitudes. Figure S2 presents different $H_{\text{k,eff}}$ profiles along with the corresponding mean amplitudes across $H_{\text{ext}}/H_{\text{k,eff}}$ from 0 to 10. When the initial drop in $H_{\text{k,eff}}$ is reduced from 5% to 2.5% [i.e., the black line to the red line in Fig. S2(a)], the optimal angle ($\theta_{\text{opt}}^{\text{Amp}}$) at which the amplitude is maximized shifts from 66° to 70°. Increasing the drop to 7.5% (the blue line) results in a peak shift to 64.5°. Varying the persisting time ($\Delta t$) between 1.00 and 2.00 ps results in only modest changes in the optimal angle (67.5° and 65°, respectively, compared to 66° for $\Delta t$ of 1.50 ps). Adjusting the recovery rate between 99.5% (faster recovery) and 98.5% (slower recovery) yields $\theta_{\text{opt}}^{\text{Amp}}$ values of 64° and 68°, respectively.

Overall, across the different profiles resulting from ultrafast demagnetization and remagnetization, the variations in $\theta_{\text{opt}}^{\text{Amp}}$ remains up to ~4°, corresponding to ~6%. These changes are insignificant in our analysis, especially given that the optimal field angle is provided as ranges. Therefore, the robustness and generality of our conclusions are maintained, despite variations in the thermal properties of the samples.

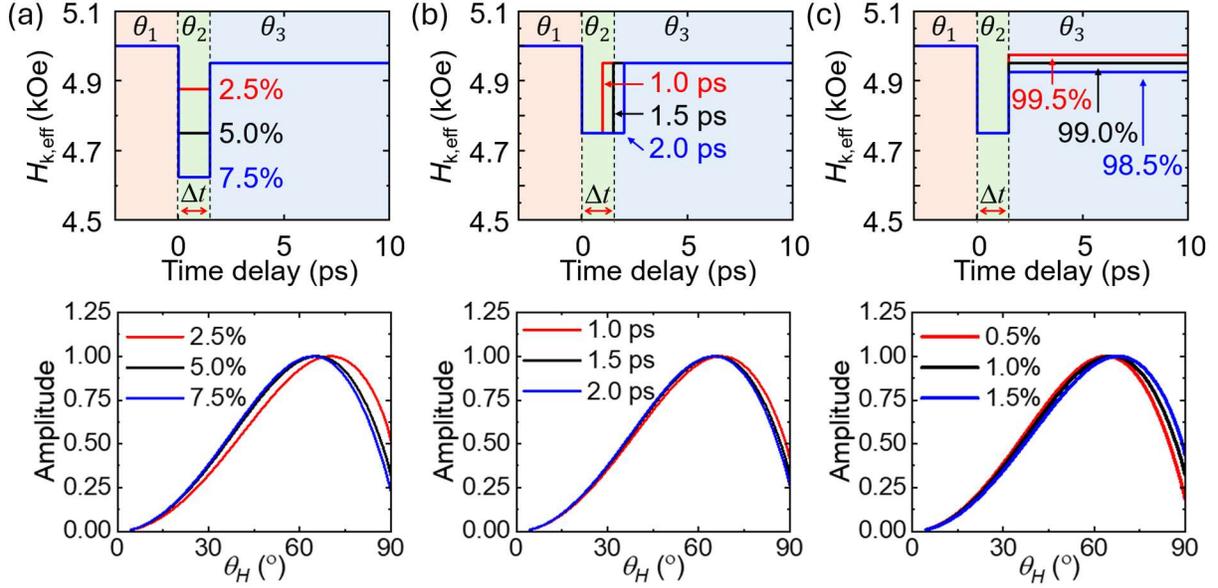

**FIG. S2.** $H_{\text{k,eff}}$ profiles (top panels) and corresponding plots of normalized mean amplitude *vs.* $\theta_H$ (bottom panels) for varitions in the initial drop rate of $H_{\text{k,eff}}$ (a), persisting time of $\Delta t$ (b), and the recovery rate of $H_{\text{k,eff}}$ (c). (a) $H_{\text{k,eff}}$ initially drops to 2.5% (red), 5.0% (black), and 7.5% (blue) of its original value, with $\Delta t$ fixed at 1.50 ps and a recovery rate of 99%. (b) $\Delta t$ is varied from 1.00 ps to 2.00 ps, while the initial drop of $H_{\text{k,eff}}$ and the recovery rate are 5% and 99%, respectively. (c) $H_{\text{k,eff}}$ recovers to 99.5%, 99%, and 98.5%, while $\Delta t$ and initial drop are 1.50 ps and 5%, respectively. Properties used in the simulation are $H_{\text{k,eff}}$ = 5.00 kOe, $\gamma$ = 18.00 rad/kOe/ns, and $H_{\text{ext}}$ ranging from 0 to 30 kOe.

The magnetization and $H_{\text{k,eff}}$ evolve during the measurement, leading to the chirp of the TR-MOKE signal. $H_{\text{k,eff}}$ will exponentially decay over time and it is simplified to be a step-function approximation in this study. While this is a simplification, it does not impact our final conclusions. To justify this simplification, we have performed additional simulations based on the Landau-Lifshitz-Gilbert (LLG) equation:[8,9]

$$\frac{d\mathbf{m}}{dt} = -\gamma \mathbf{m} \times \mathbf{H}_{\text{eff}} + \alpha \mathbf{m} \times \frac{d\mathbf{m}}{dt} \tag{S1}$$

where $\mathbf{m}$, $\gamma$, $t$, $\mathbf{H}_{\text{eff}}$, and $\alpha$ are the magnetic moment, the gyromagnetic ratio, the time, the effective magnetic field, and the Gilbert damping. In these simulations, we compare two cases: one using the step-function profile for $H_{\text{k,eff}}$, and the other using an exponentially decaying $H_{\text{k,eff}}$. The exponential profile of $H_{\text{k,eff}}$ is assumed to be proportional to the negative of rapid temperature rise. The rapid temperature rise can be captured by fitting the TR-MOKE signal with $A + Be^{-t/C} + \text{Amp} \sin(2\pi f t + \varphi)e^{-t/\tau}$, where $A$, $B$, $t$, and $C$ are the offset, the amplitude of the thermal background, the time delay, and the relaxation time of the thermal background, respectively. The third term is well explained in the manuscript. The red curve in Fig. S3(a) shows the fitted thermal background decay over time with $A = 4.61$, $B = 5.45$, and $C = 116.44$ ps. Based on this, we model the $H_{\text{k,eff}}$ decay to be proportional to the negative of the temperature rise as shown in Fig. S3(b) (red curve). This exponential profile of $H_{\text{k,eff}}$ recovers to 99.6% at 300 ps, matching the recovery rate used in our step-function simplification where $H_{\text{k,eff}}$ is restored to 99% after $\Delta t = 1.50$ ps (black line). We then simulate TR-MOKE signals using both profiles [Fig. S3(c)] and extract the resulting amplitudes (normalized to the maximal amplitude) across the range of external field angles [Fig. S3(d)]. The normalized amplitudes are obtained at $H_{\text{ext}} = 10$ kOe. In both the step-function and exponential cases, the maximum amplitude occurs at $\theta_H \approx 75^\circ$, supporting that our simplified model serves as a good approximation. This comparison demonstrates the validity of our simplification of adopting a step-function $H_{\text{k,eff}}$ profile, which does not introduce a significant discrepancy in determining $\theta_{\text{opt}}^{\text{Amp}}$, and thus does not undermine the conclusions of this study.

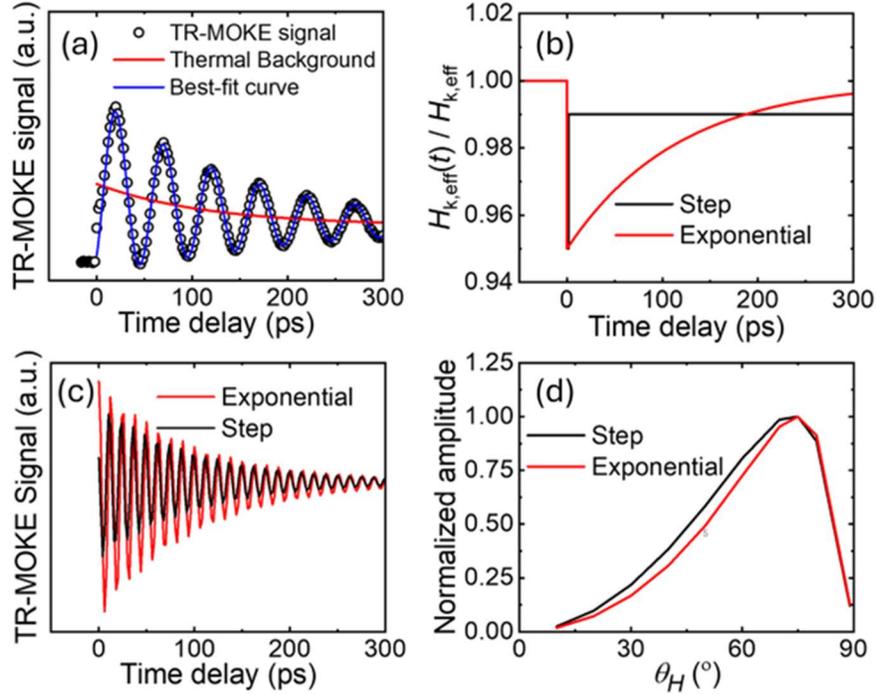

**FIG. S3.** Simulation of TR-MOKE signals with step-function and exponential $H_{k,eff}$ profiles. (a) Experimental TR-MOKE signals and the extraction of thermal background (red) with the best-fit curve (blue). (b) $H_{k,eff}$ profile with the step-function recovery (black) and the exponential recovery (red) at $\theta_H = 80°$. The exponential profile is obtained based on the assumption that $H_{k,eff}$ is proportional to the thermal background. (c) Representative simulated TR-MOKE signal based on LLG equation with step-function (black) and exponential profile (red). Properties used in the simulation are $H_{k,eff} = 5.00$ kOe, $\gamma = 18.00$ rad/kOe/ns, and $\alpha = 0.020$. $H_{ext}$ and $\theta_H$ are 10 kOe and 80°. (d) Amplitudes extracted from the simulations with different $H_{k,eff}$ profile across $\theta_H$.

## S2. Identification of $\bar{\theta}_{opt}$ based on the range of applied field

As discussed in the main article, $\bar{\theta}_{opt}^{Amp}$ depends on the range of applied fields. The normalized amplitude is calculated for various combinations of the dimensionless field ranges ($H_{ext}/H_{k,eff}$), with the lower limit ranging from 0 to 3 and the upper limit spanning from 4 to 6 (*e.g.*, 0 - 4, 0 - 5, 0 - 6, 1 - 4, 1 - 5, …, 3 - 5, and 3 - 6). Figures S4(a) and S4(b) present the normalized amplitudes calculated from the amplitude model for PMA and IMA materials, respectively. The histograms [bottom panel of Fig. S4(a)] suggest the most frequently observed optimal angles across the various field ranges. $\bar{\theta}_{\text{opt}}^{\text{Amp}}$ occurs ~64° for PMA materials and ~46° for IMA materials.

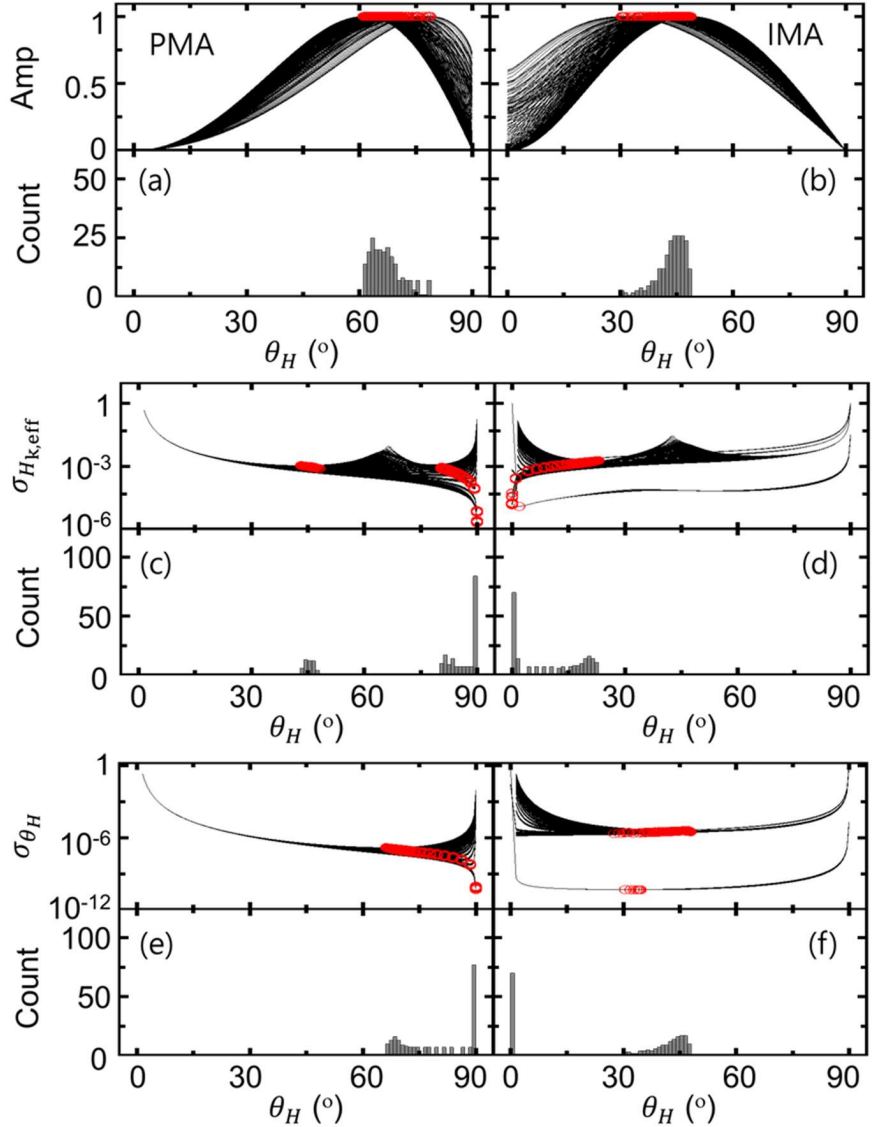

**FIG. S4.** Normalized amplitudes (upper panels) and histograms (lower panels) of $\bar{\theta}_{\text{opt}}^{\text{Amp}}$ for PMA (a) and IMA (b) materials under different field range combinations. Normalized uncertainties (upper panels) and corresponding histograms (lower panels) of $H_{\text{k,eff}}$ for PMA (c) and IMA (d) materials. Normalized uncertainties (upper panels) and corresponding histograms (lower panels) of $\theta_H$ for PMA (e) and IMA (f) materials. Red symbols denote the conditions that yield the highest amplitudes and lowest uncertainties for a given field range.

For the identification of $\bar{\theta}_{opt}^{Sx}$, the normalized uncertainties of $H_{k,eff}$ and $\theta_H$ for PMA and IMA materials are calculated and depicted in Figs. S4(c-d) and Figs. S4(e-f), respectively. Although the model calculation shows very low uncertainty to $H_{k,eff}$ and $\theta_H$ near $\theta_H = 90°$ when $H_{ext} \approx H_{k,eff}$, these conditions should be avoided in practice as magnetization precession is suppressed at 90° when $H_{ext} > H_{k,eff}$ (amplitudes are zero). For properties extracted from relaxation-rate-based analyses, the normalized uncertainties are plotted in Fig. S5. The histograms suggest that $\theta_H$ of ~66° and ~46 ° will be the optimal angle to measure $\alpha$ for PMA and IMA, respectively [Fig. S5(a) and S5(b), respectively]. It is also worth noting that the uncertainty to $\alpha$ is relatively flat with respect to $\theta_H$. As for $\sigma_{\Delta H_{k,eff}}$, it shows similar trend of $\sigma_{H_{k,eff}}$, $\theta_H$ of 90° (0 °) should be avoided for PMA (IMA) as shown in Figs. S5(c) and S5(d).

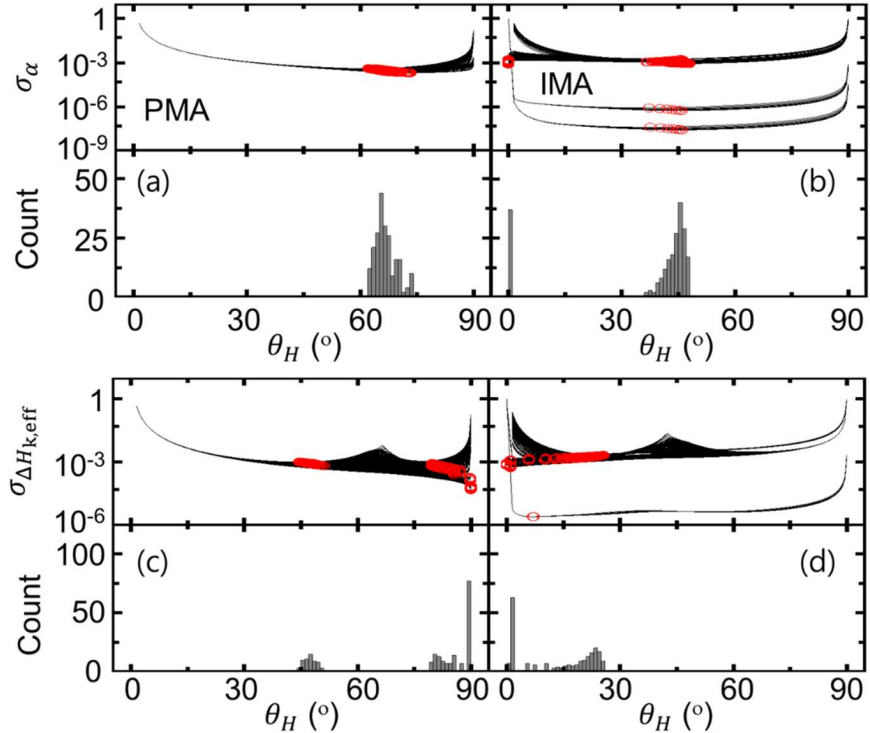

**FIG. S5.** Normalized amplitudes (upper panels) and histograms (lower panels) of $\bar{\theta}_{opt}^{\alpha}$ for PMA (a) and IMA (b) materials under different field range combinations. Normalized uncertainties (upper panels) and corresponding histograms (lower panels) of $\Delta H_{k,eff}$ for PMA (c) and IMA (d) materials. Red symbols denote the conditions that yield the lowest uncertainties for a given field range.

**S3. Sensitivities to relaxation-rate-based properties considering the easy-axis variation ($\Delta\theta_H$)**

The total relaxation rate is a summation of contributions from intrinsic damping ($\alpha$), the inhomogeneous broadening due to the variations in $H_{\text{k,eff}}$ ($\Delta H_{\text{k,eff}}$), and the inhomogeneous broadening due to the variations in the easy-axis orientation ($\Delta\theta_H$). The sensitivity to each parameter is defined as the fractional contribution of the respective relaxation rate to the total relaxation rate [Eq. (2)]. Including the $\Delta\theta_H$ term alters the total relaxation rate and thus sensitivities.

To evaluate these effects and extend the optimization model to samples with potential variations of the easy-axis orientation (*e.g.*, perpendicular $L1_0$-FePd with cubic anisotropy),[8,10] we analyze the sensitivities to relaxation-rate-based properties with the consideration of $\Delta\theta_H = 3^\circ$. Figure S6 depicts the calculated $S_\alpha$, $S_{\Delta H_{\text{k,eff}}}$, and $S_{\Delta\theta_H}$ for PMA and IMA materials as functions of the dimensionless field ($H_{\text{ext}}/H_{\text{k,eff}}$) and orientation of external fields ($\theta_H$). As shown in Figs. S6(a) and S6(b), $S_\alpha$ and $S_{\Delta H_{\text{k,eff}}}$ are reduced when the contribution from variations in the easy-axis orientation is included [as compared with the cases in Figs. 3(a) and 3(b)]. Generally, $S_\alpha$ increases and $S_{\Delta H_{\text{k,eff}}}$ decreases with $H_{\text{ext}}/H_{\text{k,eff}}$ for a fixed $\theta_H$. For $S_{\Delta\theta_H}$, it peaks at ~90° for PMA materials and ~0° for IMA materials when $H_{\text{ext}} \approx H_{\text{k,eff}}$. However, since the condition of $\theta_H \approx 90^\circ$ is typically avoided in experiments due to lack of precession, the impact of $\Delta\theta_H$ is suppressed as well.

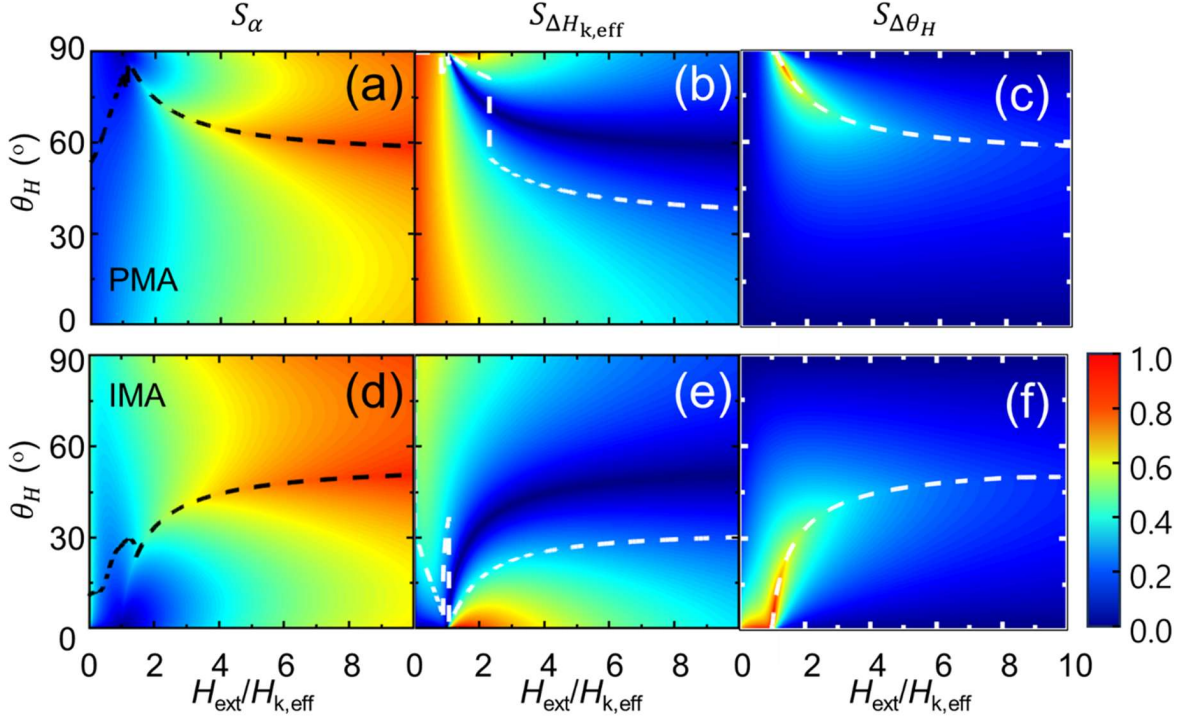

**FIG. S6.** Contour plots of sensitivities to $\alpha$ [(a) and (d)], $\Delta H_{k,\text{eff}}$ [(b) and (e)], and $\Delta\theta_H$ [(c) and (f)] of PMA (a-c) and IMA (d-f) materials under varying $H_{\text{ext}}/H_{k,\text{eff}}$ and $\theta_H$ with consideration of $\Delta\theta_H = 3°$. Black and white curves represent the recommended $\theta_{\text{opt}}^{S_X}$ at each $H_{\text{ext}}/H_{k,\text{eff}}$ to achieve the optimal sensitivity.

## S4. Amplitude compensation using a Soleil-Babinet compensator

A custom-designed sample holder in the Voigt geometry[11] is used to enable low-$\theta_H$ measurements with a gold mirror [Fig. S7(a)]. When the reflected probe beam carries a polarization rotation due to the Kerr effect, it reflects off the gold mirror and experiences a phase shift. This phase shift affects the TR-MOKE signal amplitude and must be compensated. The reflection phase due to the gold mirror varies with the angle of incidence [AOI in Fig. S7(a)] and can be compensated by using a Soleil-Babinet compensator (ThorLabs, SBC-VIS) which allows fine control *via* a micrometer. TR-MOKE signal amplitudes are measured as a function of the micrometer displacement ($x$) at a given AOI and fitted to $a\sqrt{(\sin^2(bx + c))}$ with $a$ being the amplitude, $b$ being the frequency of compensator-induced modulation, and $c$ being the offset [Fig. S7(b)]. From the fit, we can determine both the displacement-to-phase retardation conversion factor and phase correction required for accurate compensation of TR-MOKE signals due to the incorporation of the gold mirror at any arbitrary AOI.

To validate this approach, TR-MOKE amplitudes are measured at different AOI conditions of 65° and 75°, which correspond to $\theta_H$ = 40° and 60° [red and blue symbols in Fig. S7(c), respectively]. These TR-MOKE amplitudes are compared with one measured at $\theta_H$ = 80° with a regular sample holder (no gold mirror or compensator, green symbol). With phase correction using the compensator, TR-MOKE signal amplitudes are consistent for varying $\theta_H$, deviating by less than ~7% despite the use of different AOI conditions and sample holders.

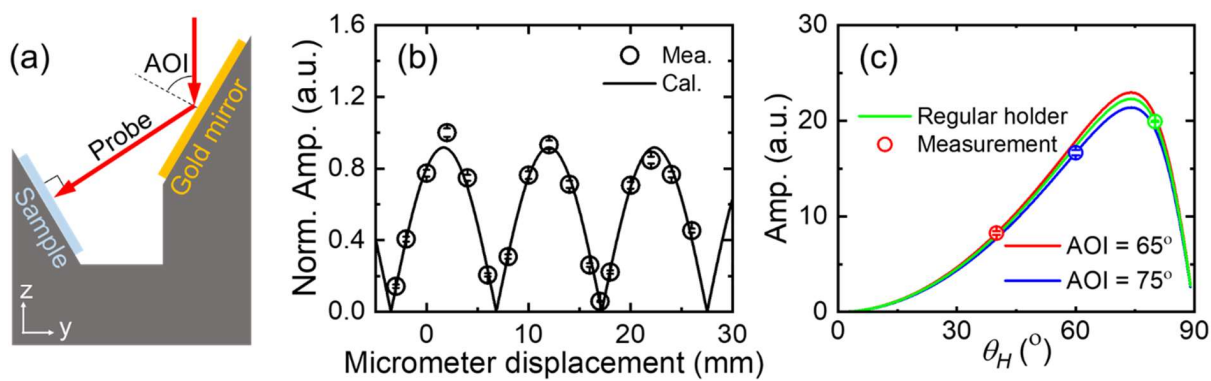

**FIG. S7**. (a) Schematic of the custom-designed sample holder in the Voigt geometry. AOI denotes the angle of incidence. (b) Normalized TR-MOKE signal amplitudes as a function of the micrometer displacement and its best fit at 10 kOe of $H_{ext}$. (c) Normalized TR-MOKE signal amplitudes with 10 kOe of $H_{ext}$ for different AOI conditions: measurements with the gold mirror and compensator at AOI = 65° (red symbol) and 75° (blue symbol) closely match results from a regular sample holder without a gold mirror or a compensator (green symbol), demonstrating effective amplitude compensation for the use of the Voigt sample holder with a gold mirror. Lines are from calculations based on Eq. (1).

## S5. TR-MOKE data reduction for property extraction

Based on Kittel's model, the frequency can be calculated:[12]

$$f = \gamma / 2\pi \sqrt{H_1 H_2} \tag{S2}$$

$$H_1 = H_{ext} \cos(\theta_e - \theta_H) + H_{k,eff} \cos^2\theta_e \tag{S3}$$

$$H_2 = H_{ext} \cos(\theta_e - \theta_H) + H_{k,eff} \cos(2\theta_e) \tag{S4}$$

The equilibrium angle, $\theta_e$, can be calculated using the following equation by minimizing the magnetic free energy:

$$2H_{ext} \sin(\theta_H - \theta_e) = H_{k,eff} \sin(2\theta_e). \tag{S5}$$

The field-dependent $f$ is measured by sweeping $H_{ext}$ in our TR-MOKE measurements. By fitting the field-dependent $f$ extracted from TR-MOKE signal ($\Delta\theta_k$) to Eq. (S2), $\gamma$, $H_{k,eff}$, and $\theta_H$ can be determined as shown in Fig. S8. Both frequency and properties are extracted through least-square fitting.[13] In the case of the relaxation-time-related properties (i.e., $\alpha$, $\Delta H_{k,eff}$, and $\Delta\theta_H$), the field-dependent $1/\tau$ is measured by sweeping $H_{ext}$ and fitted to Eq. (2).

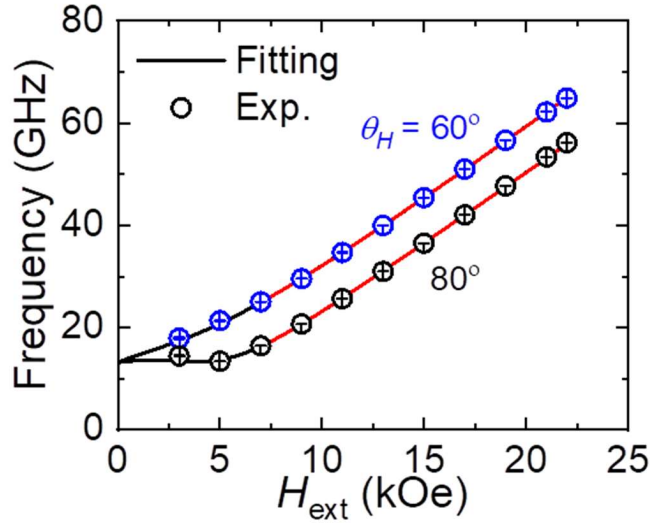

**FIG. S8.** Field-dependent frequency when $\theta_H = 60°$ (blue symbols) and 80° (black symbols). Black curves represent the best-fit curves and red curves represent the best-fit curves within the fitting window (7 to 22 kOe).

## S6. Optimization recommendation for the sweeping-angle configuration

In the case where $H_{ext}$ is fixed and $\theta_H$ is varied (scenario 2, the sweeping-angle configuration), the optimized field ($H_{opt}$) can be determined in a similar manner as $\theta_{opt}$ in the sweeping-field configuration (scenario 1). At each angle, $H_{opt}$ is chosen as the $H_{ext}$ that exhibits the highest amplitude or the largest value of the objective function ($S_X \times$ Amp), where $S_X$ is the measurement sensitivity to a specific parameter. Regarding the amplitude, $H_{opt}/H_{k,eff}$ is ~1 when $\theta_H$ is 90° for PMA materials [Fig. S9(a)]. As $\theta_H$ decreases to 0°, $H_{opt}/H_{k,eff}$ becomes higher and saturates at $H_{opt}/H_{k,eff} \approx 7.5$. Conversely, for IMA materials [Fig. S9(d)], $H_{opt}/H_{k,eff}$ starts at ~1 for $\theta_H = 0°$ and increases as $\theta_H$ approaches 90°, saturating ~6. For optimizing sensitivity to $H_{k,eff}$ and $\theta_H$, the optimization model suggests that $H_{opt}/H_{k,eff}$ is ~1 for both PMA [Figs. S9(b) and S9(c)] and IMA [Figs. S9(e) and S9(f)] materials. However, similar to the sweeping-field case, $H_{ext}/H_{k,eff} \approx 1$ should be avoided due to incoherent precession.

Figure S10 depicts $H_{opt}/H_{k,eff}$ for optimizing the sensitivities to $\alpha$ and $\Delta H_{k,eff}$. For PMA materials, $H_{opt}/H_{k,eff}$ for $\alpha$ increases from 1 to 10 as $\theta_H$ decreases from 90° to 0° [Fig. S10(a)]. For IMA materials [Fig. S10(c)], $H_{opt}/H_{k,eff}$ for $\alpha$ is 1 when $\theta_H$ is 0° and increases with $\theta_H$. $H_{opt}/H_{k,eff}$ for $\Delta H_{k,eff}$ remains near 1 for both PMA and IMA materials [Figs. S10(b) and S10(d)]. Again, to ensure coherent precession with sufficient amplitudes, it is recommended to avoid operating TR-MOKE measurements exactly at this field ratio.

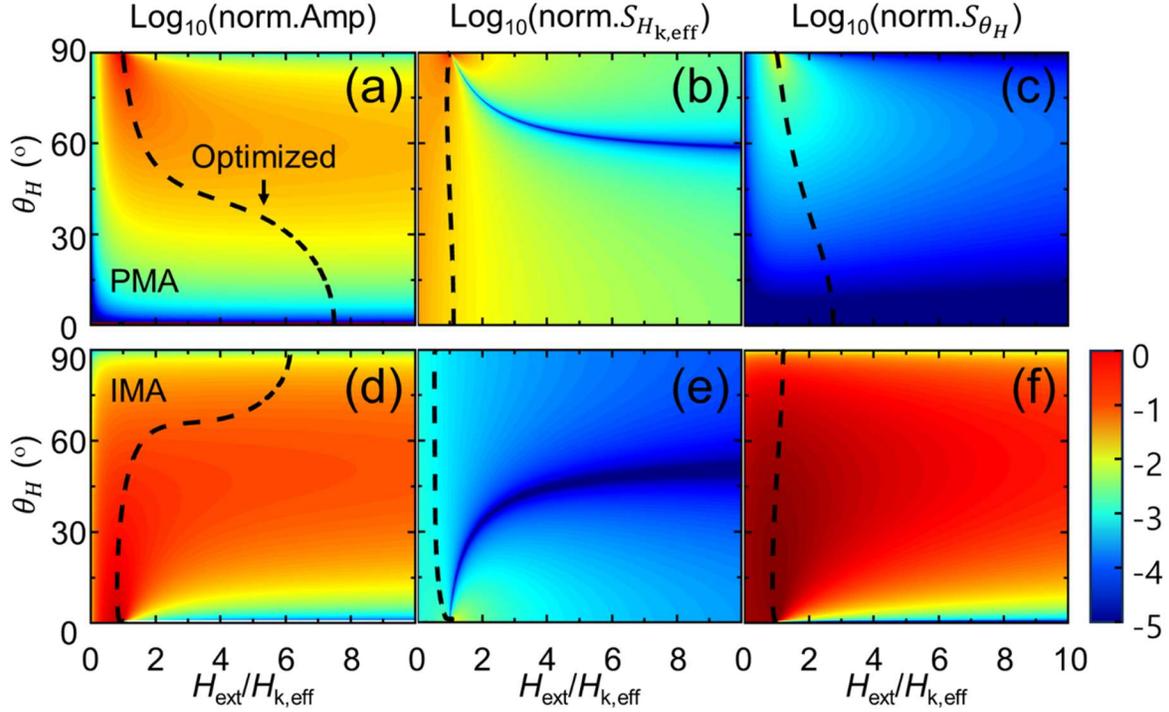

**FIG. S9.** Contour plots of normalized amplitudes [(a) and (d)], sensitivities to $H_{\text{k,eff}}$ [(b) and (e)] and sensitivities to $\theta_H$ [(c) and (f)] of PMA [(a) – (c)] and IMA [(d) – (f)] materials under varying $H_{\text{ext}}/H_{\text{k,eff}}$ and $\theta_H$ (all in the logarithmic scale). Black dashed curves represent $H_{\text{opt}}$ as a function of $\theta_H$ for the highest amplitude and maximum value of the objective function for the sweeping-angle configuration. Parameters used are $H_{\text{k,eff}} = \pm 5.00$ kOe, $\Delta H_{\text{k,eff}} = 0.50$ kOe, $\gamma = 18.00$ rad/kOe/ns, and $\alpha = 0.020$, close to the magnetic properties of the CoFeB reference sample used in sweeping-field experimental validation. Color bar ticks are negative as data are normalized and $\log_{10}$-scaled.

Similar to scenario 1 in the main article, the optimal field shall be recommended based on the range of applied field angles ($\theta_H$) for practical implementation in measurements. The starting angle ranges from 0° to 30° and the final angle spans from 60° to 90°. Figures S11(a) and S11(b) show the normalized amplitudes with different combinations of angle ranges and the corresponding histograms identifying the most frequently occurring optimal fields. The same approach is used to assess uncertainties of the frequency-based properties, including $H_{\text{k,eff}}$ [Figs. S11(c) and S11(d)] and the field angle [Figs. S11(e) and S11(f)]. For frequency-based properties, the optimized fields for maximizing amplitudes and sensitivities are concentrated on $H_{\text{ext}}/H_{\text{k,eff}} \approx 1$ for both PMA and IMA materials as shown in Fig. S11. In the case of relaxation-rate-based properties, the most frequently observed optimal fields with lowest uncertainties are $H_{ext}/H_{k,eff} \approx 1$ as well (Fig. S12). It is worth noting that the optimal fields for $\alpha$ spread widely for PMA, not being sensitive to the external field for $H_{ext} > H_{k,eff}$ in the sweeping-angle configuration. Considering the dispersion in $H_{k,eff}$, a magnetic domain with $H_{k,eff} + \Delta H_{k,eff}$ will not saturate at $H_{ext} = H_{k,eff}$, leading to incoherent precession at $H_{ext} \approx H_{k,eff}$. Therefore, it is recommended to use a field higher than $H_{k,eff} + \Delta H_{k,eff}$ to promote coherent precession. Due to the challenge in direct estimation of $\Delta H_{k,eff}$, a practical recommendation is to use $H_{ext}/H_{k,eff} \approx 2$ for coherent precession, reasonably high amplitudes, and relatively good sensitivities.

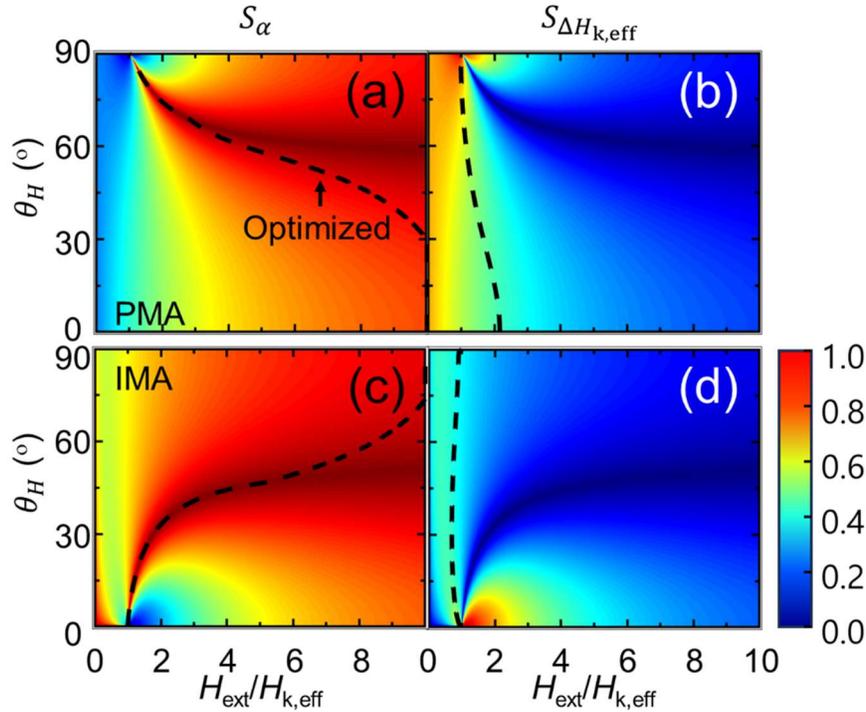

**FIG. S10.** Contour plots of sensitivity to $\alpha$ [(a) and (c)] and $\Delta H_{k,eff}$ [(b) and (d)] for PMA [(a) and (b)] and IMA [(c) and (d)] materials under varying $H_{ext}/H_{k,eff}$ and $\theta_H$. Black curves represent the recommended $H_{opt}/H_{k,eff}$ at each $\theta_H$ to achieve optimal sensitivity.

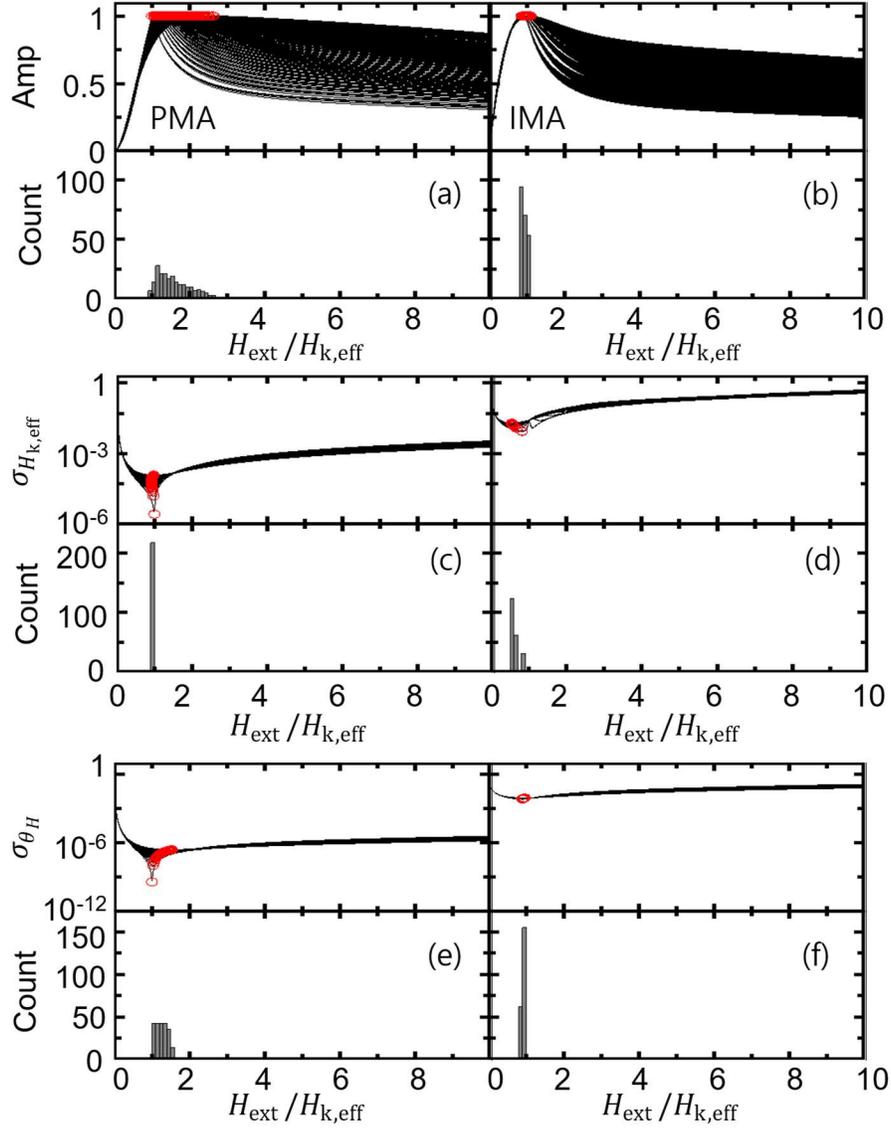

**FIG. S11.** Normalized amplitudes [upper panels, (a) and (b)] and histograms for the fixed optimal field (bottom panels) of PMA (a) and IMA (b) materials for sweeping-angle configuration. Normalized uncertainty of $H_{k,eff}$ [(c) and (d)] and $\theta_H$ [(e) and (f)] of PMA [(c) and (e)] and IMA [(d) and (f)] materials and the respective histograms. Red symbols represent the highest amplitudes and lowest uncertainties at a given field range.

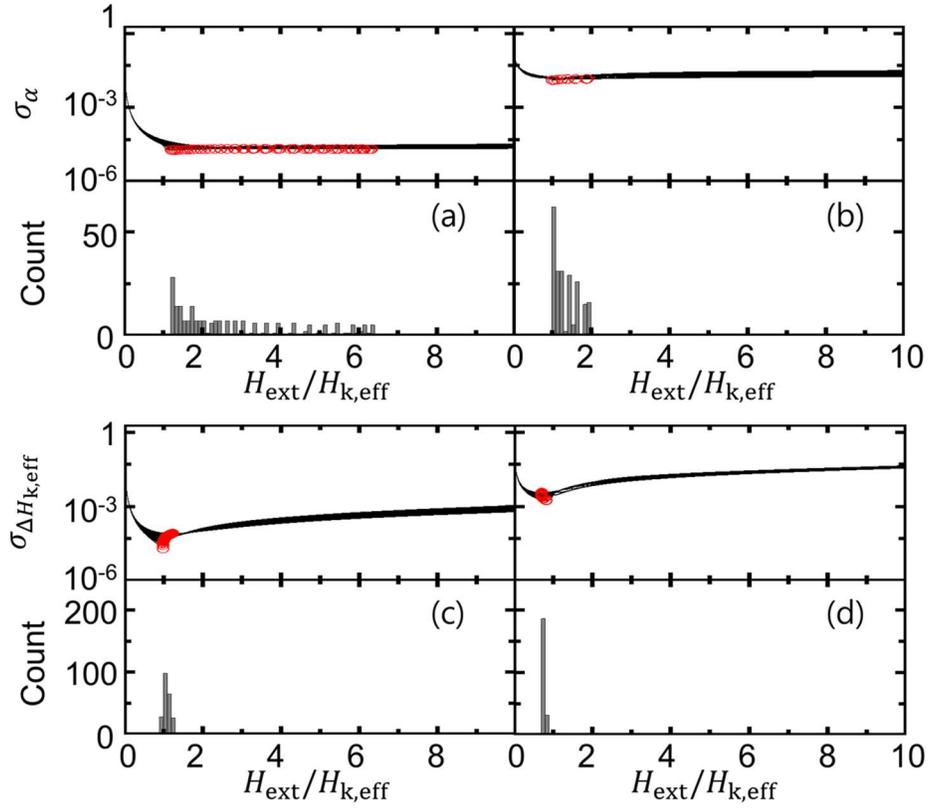

**FIG. S12.** Normalized uncertainty of $\alpha$ [upper panels, (a) and (b)] and $\Delta H_{\mathrm{k,eff}}$ [bottom panels, (c) and (d)] of PMA [(a) and (c)] and IMA [(b) and (d)] materials and the respective histograms for sweeping-angle configuration. Red symbols represent the lowest uncertainties at a given field range.